\documentclass{moriond}

\usepackage{siunitx}
\usepackage{amssymb} 
\usepackage{lineno}
\usepackage{multirow}
 
\def\be{\begin{equation}}
\def\ee{\end{equation}}
\def\bea{\begin{eqnarray}}
\def\eea{\end{eqnarray}}

\newcommand{\Photo}{\includegraphics[height=35mm]{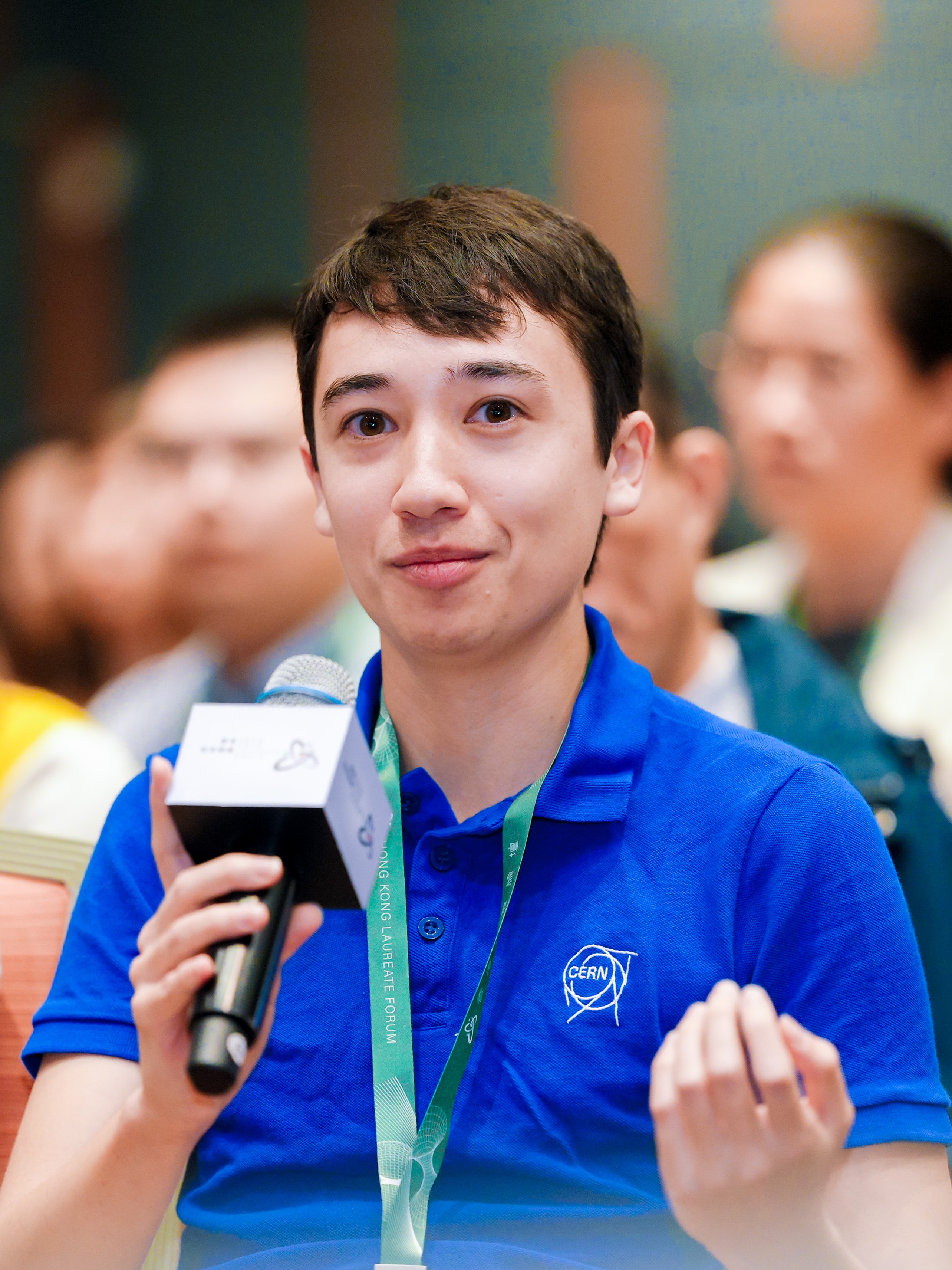}}

\begin{document}
\vspace*{4cm}
\title{ATLAS wildcard:\\Searches for massive, long-lived particles in events with displaced vertices with ATLAS}

\author{ David Rousso\\on behalf of the ATLAS Collaboration\footnote{Copyright 2026 CERN for the benefit of the ATLAS Collaboration. CC-BY-4.0 license.}}

\address{Deutsches Elektronen-Synchrotron (DESY)\\Notkestraße 85, 22607 Hamburg, Germany}

\maketitle\abstracts{
Many recent efforts at the LHC have been made to search for new particles that do not decay promptly but are instead long-lived. 
This has been done via many different exotic signatures, including searches performed at ATLAS for displaced vertices (DV), where the new long-lived particle decays into multiple visible tracks after having traveled a certain distance into the detector. 
This talk covers two such searches: a Run 2 search for DVs in events triggered by missing transverse energy, and a Run 3 search for DVs in events triggered by muons. 
The former search is the first to use a new "fuzzy" displaced vertex reconstruction algorithm, alongside the standard one, to effectively reconstruct cases where the long-lived particle decays into heavy quarks that are themselves slightly long-lived, hence causing the final visible decay products to not point back exactly to the same vertex, setting limits on Higgs Portal, SUSY, and DFSZ axino models. 
The latter search is the first to use a new displaced muon trigger, setting limits on RPV SUSY models.}

\section{An Introduction to Long-Lived Particles and Displaced Vertices}
The solutions to some of the major problems of the Standard Model (SM) usually involve new particles.
These new particles beyond the SM (BSM) that are searched for in colliders can either decay immediately after they are produced, decay while they are traveling through the detector, or decay after passing through the whole detector, if at all\cite{DetectorPaper}. 
The majority of analyses in ATLAS at the LHC focus on the "prompt" cases, where the particle decays immediately within $\tau \lesssim \SI{0.0033}{ns}$ or $c\tau \lesssim \SI{1}{mm}$, as this is what the reconstruction algorithms and detectors have been optimized for.
However, there is no a-priori theoretical motivation for why new particles must necessarily decay this quickly. In fact several well-motivated mechanisms can cause new particles to be "long-lived" (LLPs), making it important to search for these special cases.

LLPs can produce a variety of possible exotic detector signatures depending on their nature. 
One of which, the displaced vertex (DV), is produced when an LLP decays into many charged particles (usually hadronic) while traveling through the inner tracker detector. 
Tracks made by these charged decay products would point back to a vertex that is displaced from the original proton-proton interaction.
To reconstruct these displaced tracks, a dedicated pass of a special tracking algorithm, known as Large-Radius Tracking (LRT), is required as the standard tracking algorithm constrains tracks it reconstructs to not be displaced due to computational considerations.
Various algorithms exist to reconstruct the DVs from these displaced tracks, usually requiring that all tracks must be consistent with pointing back to the exact same point.
DVs themselves can be characterized by their number of reconstructed tracks ($N_{\textrm{tracks}}$), combined invariant mass of its tracks ($m_{\textrm{DV}}$), and the position of the vertex ($\textrm{DV}_{rxy/x/y/z}$).

DVs are challenging to trigger on directly when deciding which collision events to save for analysis. 
As muons, jets, or missing transverse energy (MET) are easier to trigger on, a common strategy is to search for DVs in events triggered by such objects.
The two such analyses presented here search for DVs in events that have been triggered by MET\cite{DVMET} and a muon\cite{DVMuon} respectively.

\section{Search for Displaced Vertices in Events Triggered by Missing Transverse Energy}
This full Run 2 (2016-2018, \SI{137}{fb^{-1}}) analysis\cite{DVMET} is a follow-up of the partial Run 2 (2016-only, \SI{32.8}{fb^{-1}}) search\cite{DVMETOLD}. 
On top of the increase in data, improvements from the previous search include the introduction of a signal region searching for two or more DVs in the event as well as the introduction of fuzzy vertexing.
As the chances of finding two or more background DVs in the event are much lower than finding just one, requiring two or more of them allows the relaxation of requirements on the number of tracks or mass, resulting in higher sensitivity to lighter LLPs. 

\begin{figure}
\centering
\begin{minipage}{0.45\linewidth}
\centerline{\includegraphics[width=\linewidth]{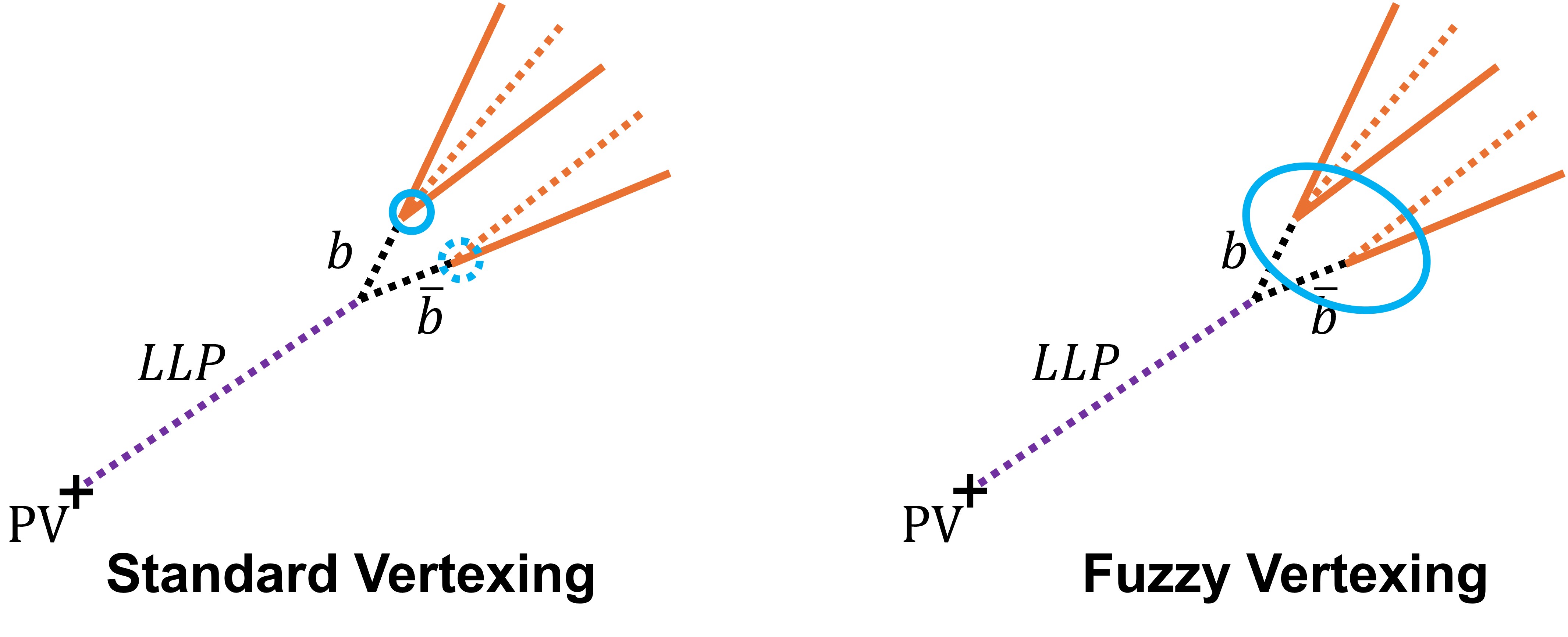}}
\end{minipage}
\hspace{0.05\linewidth}
\begin{minipage}{0.40\linewidth}
\centerline{\includegraphics[width=\linewidth]{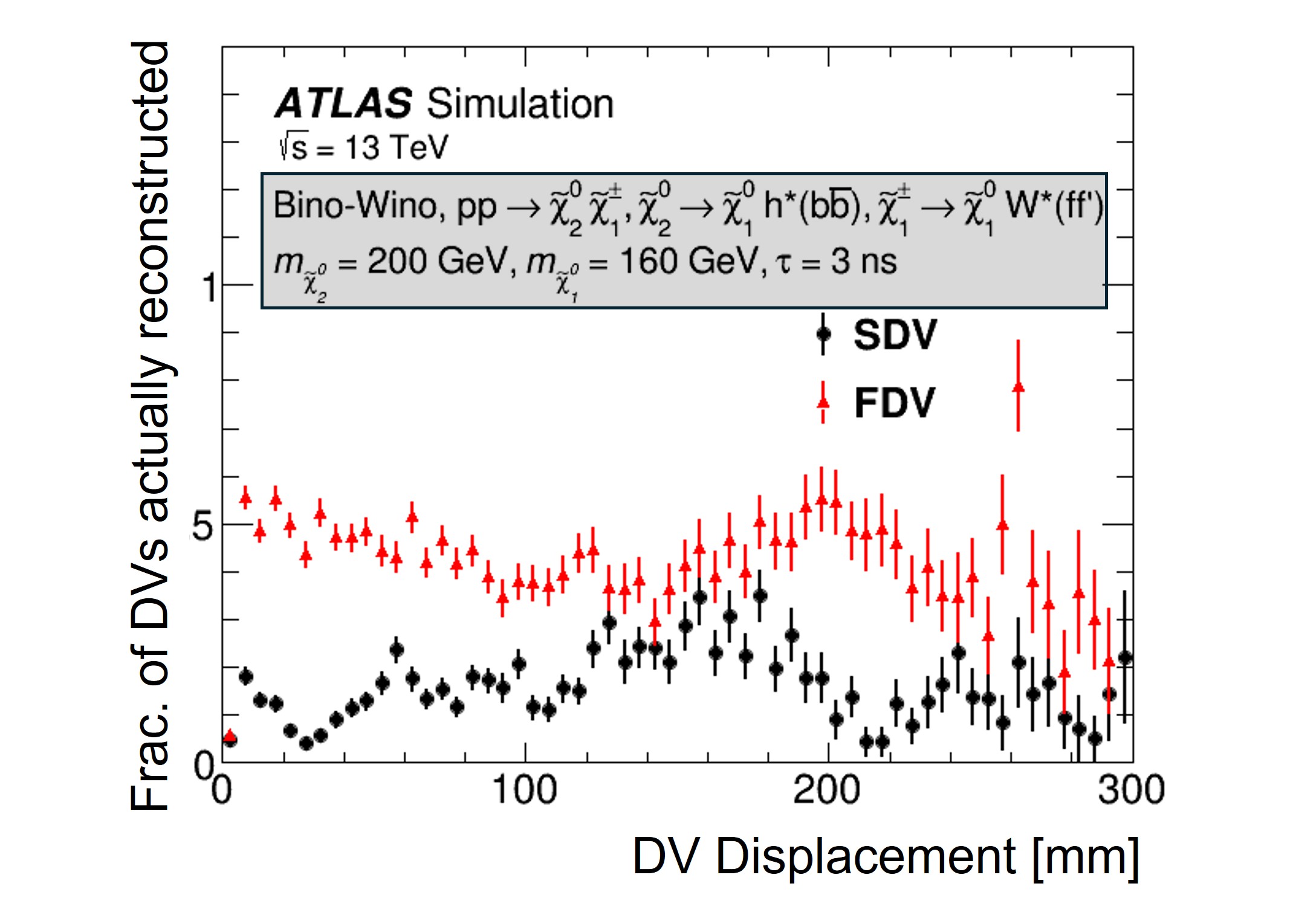}}
\end{minipage}
\caption[]{Left: Schematic of heavy flavour LLP decays in standard vertexing vs. fuzzy vertexing, with the vertices themselves circled in blue. Right: Vertexing efficiency of a simulated signal with heavy flavour decays, comparing standard vs. fuzzy vertexing\cite{DVMET}.}
\label{fig:fuzzy}
\end{figure}

If an LLP decays into heavy quarks that are themselves slightly long-lived, then the tracks of the final decay products of the LLPs will no longer point back to the exact same point.
As the reconstruction efficiency of such displaced tracks is not perfect, chances of reconstructing the subvertices of the heavy quark decays individually are very low, and thus risk not reconstructing anything at all for these cases. 
The solution is to relax the requirement that all tracks must point back to the exact same point, where one allows the vertex to be a fuzzy volume instead, as shown in Figure~\ref{fig:fuzzy} (left).
This drastically increases the reconstruction efficiency for LLPs that decay into heavy quarks, as can be seen in Figure~\ref{fig:fuzzy} (right). 

\subsection{Selections}
All events must pass the MET triggers and have MET $\geq \SI{150}{GeV}$. 
There is no requirement that the MET direction needs to be aligned with the DV in any way.

The analysis is divided into 3 signal regions (SRs): 1 or more Standard DV (1SDV), 1 Fuzzy DV (1FDV), and 2 or more Fuzzy DVs (2FDV). 
"Standard DV" denotes a DV reconstructed with standard vertexing, while "Fuzzy DV" denotes a DV reconstructed with the new fuzzy vertexing algorithm.
This allows for the targeting of various scenarios and models.
The 1SDV and 1FDV SRs require DVs have 5 or more tracks and an invariant mass of $m_{\textrm{DV}}\geq \SI{10}{GeV}$. 
The 2FDV SR allows the requirement to be relaxed to 4 or more tracks and $m_{\textrm{DV}}\geq\SI{1.5}{GeV}$.
The fuzzy DV SRs are combined when interpreting as they are orthogonal, however the standard vertexing SR is kept effectively as a separate analysis.

DVs must not be located inside detector material, must lie within $|\textrm{DV}_{rxy}|\leq\SI{300}{mm}$ and $|\textrm{DV}_{z}|\leq\SI{300}{mm}$, and have more than \SI{4}{mm} transverse distance away from any collision vertex.

\subsection{Background Estimation}
Background processes that produce DVs include normal particles interacting hadronically with detector material, as well as standard model (SM) LLPs such as $B$-hadrons or $K_s^0$. 
Note that these processes are correlated to the number of tracks and jets in the event and are completely independent from MET, muons, or photons in the event.

As no signals considered in the search are associated with photons, and photons are not correlated with DV activity, the photon-triggered region is used as a control region.
To simplify, the fraction of events that have DVs, accounting for the track and jet activity, is therefore determined in this photon-triggered region, and multiplied by the total number of events in the MET-triggered region given the track and jet activity to obtain the estimated number of DVs in the MET-triggered region. 
Validation regions are made by inverting the requirements that DVs must not be in detector material, have a certain number of tracks, mass, or seeds in the vertexing stage.

\subsection{Results}
The number of estimated background events in the different signal regions and relative contributions from the various sources of uncertainties, as well as the actual number of observed events is shown in Table~\ref{tab:DVMETResult}. The dominant uncertainty is statistical. No significant excess was observed.

\begin{table}
\caption[]{Number of observed and expected events in the various signal regions, as well as the relative contributions of the different sources of uncertainties\cite{DVMET}.}
\label{tab:DVMETResult}
\vspace{0.4cm}
\begin{center}
\begin{tabular}{l|c|c|ccc}
\hline
\multirow{2}{*}{Region} & \multirow{2}{*}{Observed Evnts.} & \multirow{2}{*}{Estimated Bkg.} & \multicolumn{3}{c}{Uncertainties}    \\
                        &                                      &                                       & Statistical & Pileup   & Non-closure \\ \hline
$\geq$ 1 Standard DV          & 1                                    & $0.6 \pm 0.4$                         & $\pm 70\%$   & $\pm 5\%$ & -           \\ \hline
1 Fuzzy DV              & 3                                    & $0.8 \pm 0.5$                         & $\pm 57\%$   & $\pm 2\%$ & $\pm 22\%$   \\
$\geq$ 2 Fuzzy DVs            & 2                                    & $1.3 \pm 0.6$                         & $\pm 45\%$   & $\pm 4\%$ & $\pm 20\%$   \\ \hline
\end{tabular}
\end{center}
\end{table}

\begin{figure}
\begin{minipage}{0.9\linewidth}
\centerline{\includegraphics[width=\linewidth]{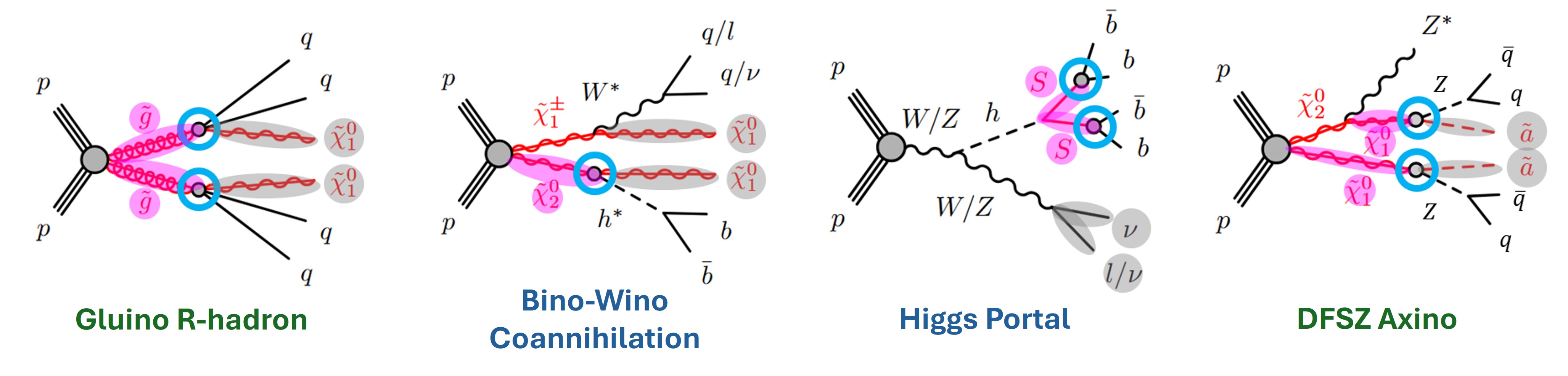}}
\end{minipage}
\caption[]{Example Feynman diagrams of the benchmark models used\cite{DVMET}. BSM particles are in red. LLPs are highlighted in pink, processes causing MET are highlighted in grey, and the DV is circled in blue.}
\label{fig:DVMETFeynman}
\end{figure}

Four benchmark models were used with various LLP decay modes and number of LLPs to target the various signal regions. 
These are the gluino $R$-hadron\cite{RHadron}, Wino-Bino co-annihilation\cite{WinoBino}, Higgs portal\cite{HiggsPortal}, and Dine-Fischler-Srednicki-Zhitnitsky (DFSZ) axino models\cite{axino}, with example Feynman diagrams shown in Figure~\ref{fig:DVMETFeynman}.
The $R$-hadron gluino and DFSZ axino models are targeted the 1SDV SR with its light LLP decays, whereas the Wino-Bino and Higgs portal are targeted the 1FDV and 2FDV SRs with its one and two heavy LLP decays respectively. 

\begin{figure}[h]
\begin{minipage}{0.5\linewidth}
\centerline{\includegraphics[height=0.6\linewidth]{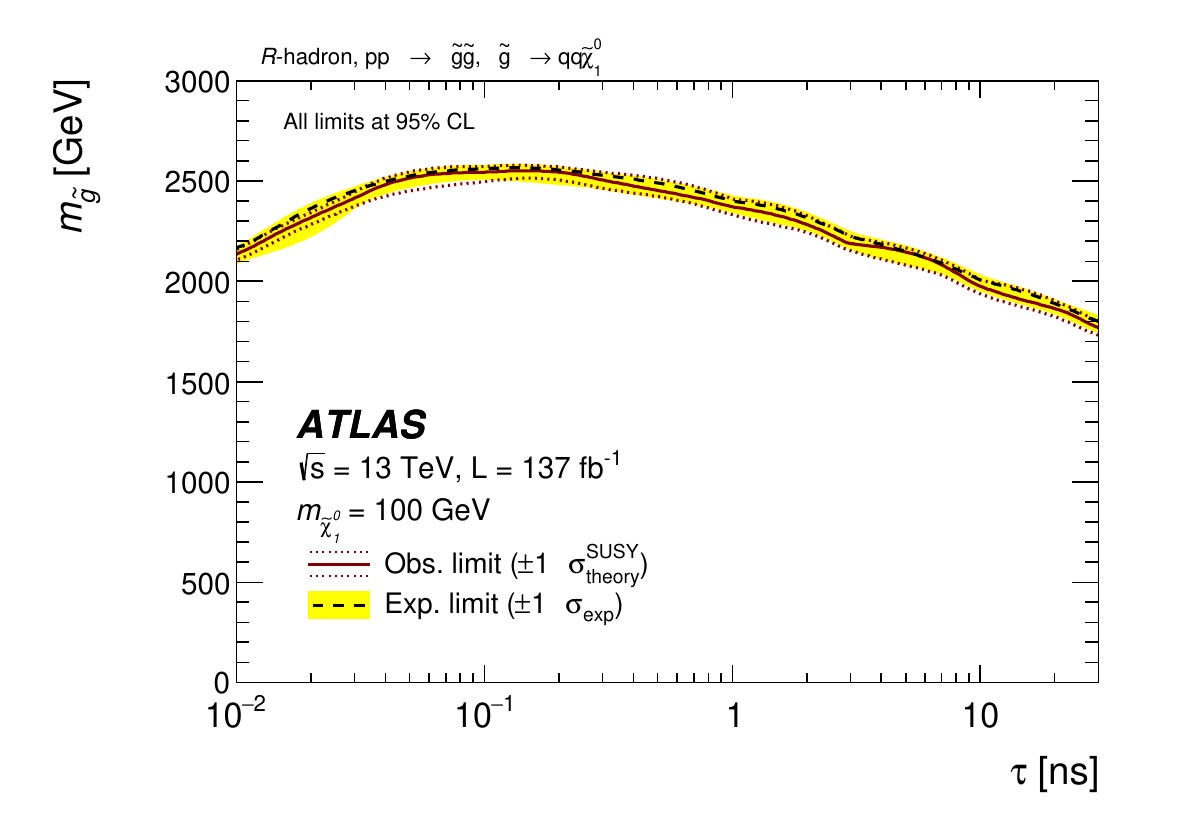}}
\end{minipage}
\hfill
\begin{minipage}{0.5\linewidth}
\centerline{\includegraphics[height=0.6\linewidth]{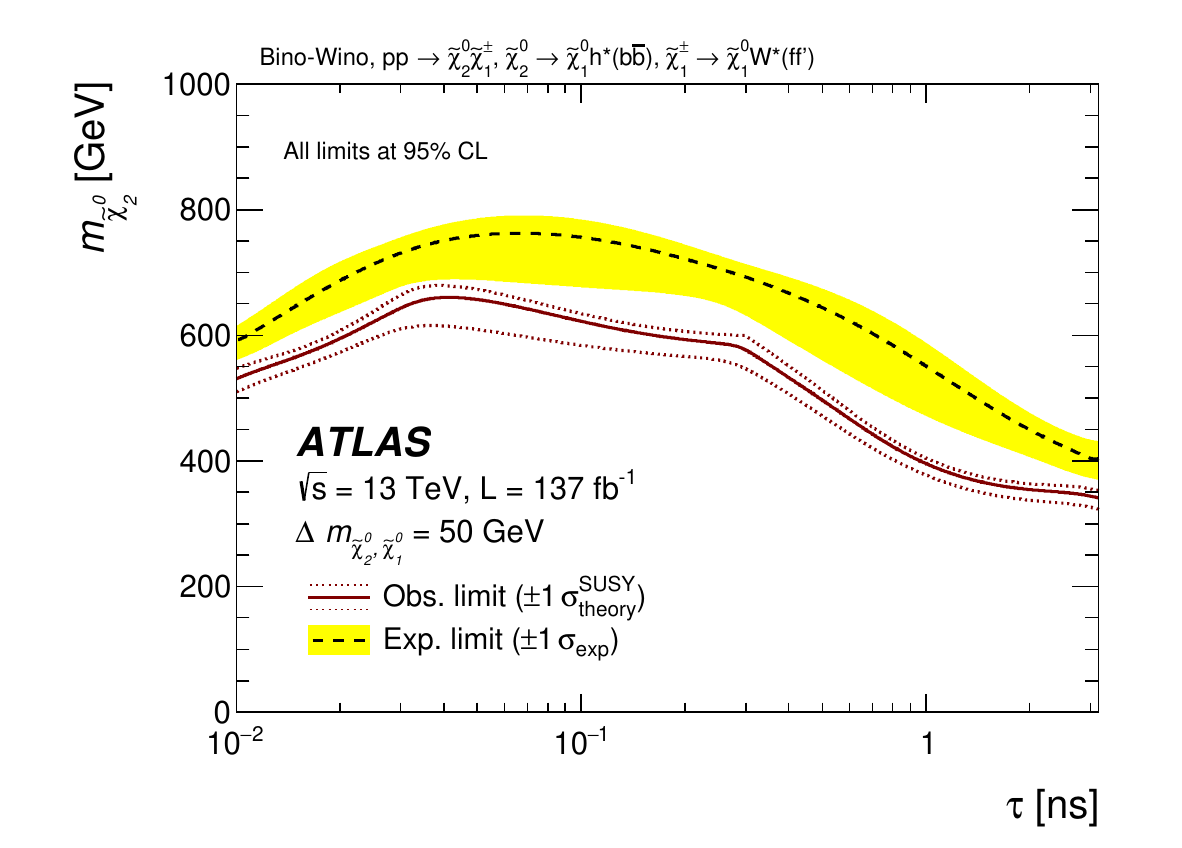}}
\end{minipage}
\begin{minipage}{0.5\linewidth}
\centerline{\includegraphics[height=0.5\linewidth]{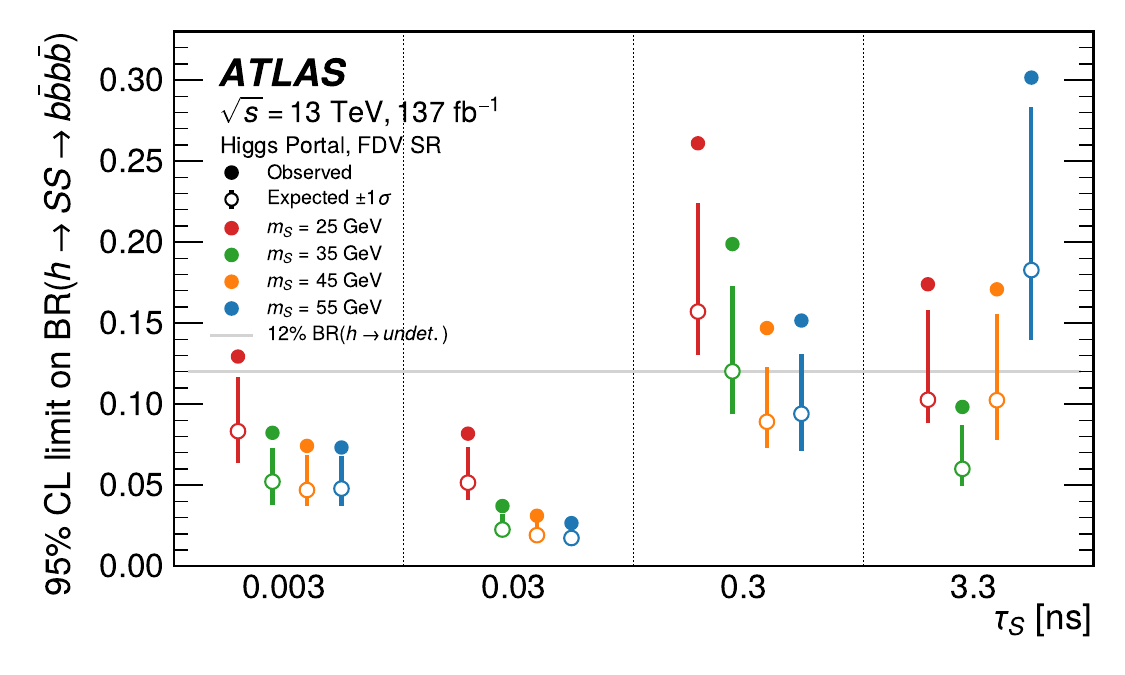}}
\end{minipage}
\hfill
\begin{minipage}{0.5\linewidth}
\centerline{\includegraphics[height=0.57\linewidth]{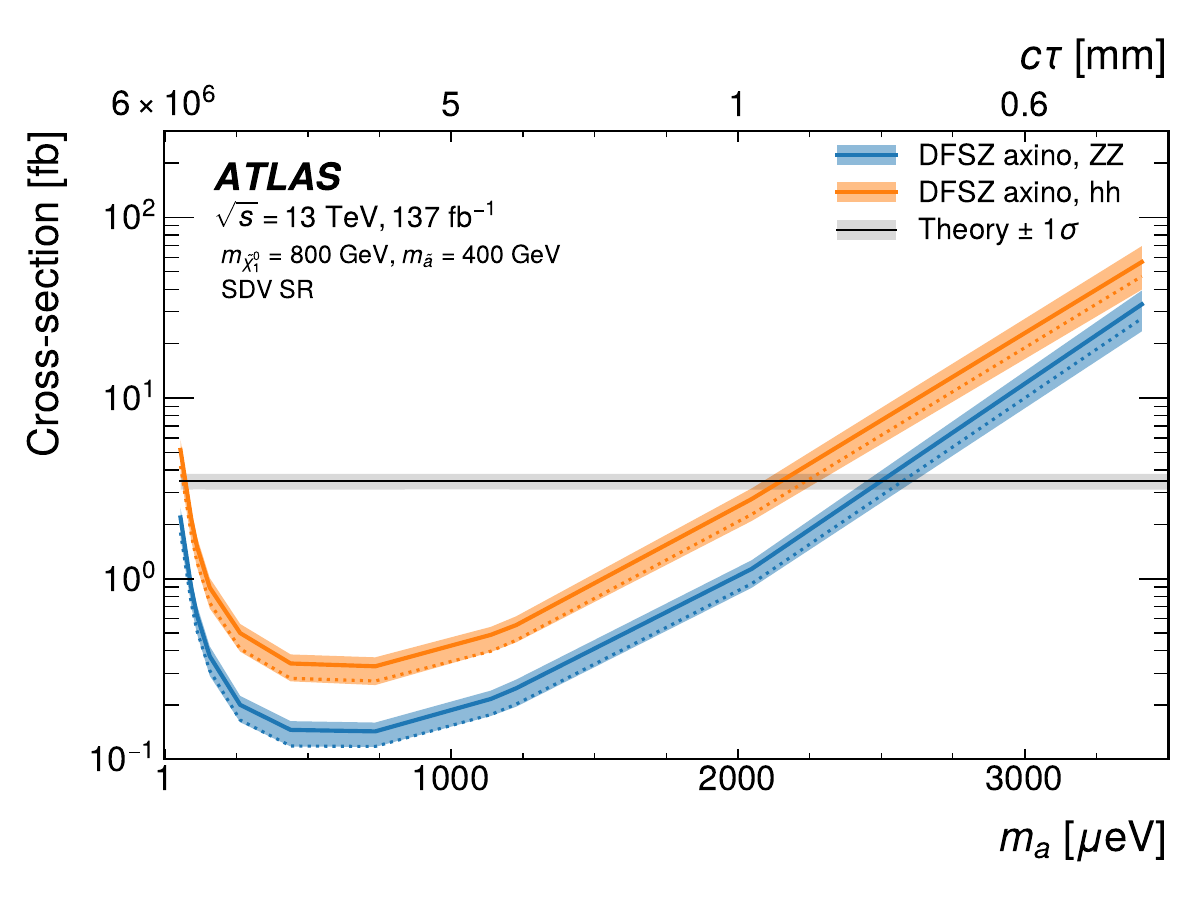}}
\end{minipage}
\caption[]{Example exclusion limits for the gluino R-hadron, Bino-Wino co-annihilation, Higgs portal, and DFSZ axino models respectively\cite{DVMET}.}
\label{fig:DVMETLimits}
\end{figure}

Selected exclusion limits from each model are shown in Figure~\ref{fig:DVMETLimits}. 
The gluino $R$-hadron limits represent an approximate 200 to \SI{300}{GeV} increase in exclusion in the peak regions with respect to the previous analysis\cite{DVMETOLD}.
The Wino-Bino results have now been included in the SUSY summary plots\cite{SUSYSummaryPlots}.
The Higgs portal model has high sensitivity at the high lifetimes as this represents the case where one of the LLPs has decayed outside of the detector and has been reconstructed as MET, meaning that the neutrinoless but relatively high cross-section gluon gluon fusion Higgs production mode now contributes to the sensitivity.
The DFSZ axino limits represent the first explicit limits on this scenario, and limits can be set on the axion itself as well.

\section{Search for Displaced Vertices in Events Triggered by Muons}
\begin{figure}[h!]
\centering
\begin{minipage}{0.3\linewidth}
\centerline{\includegraphics[width=\linewidth]{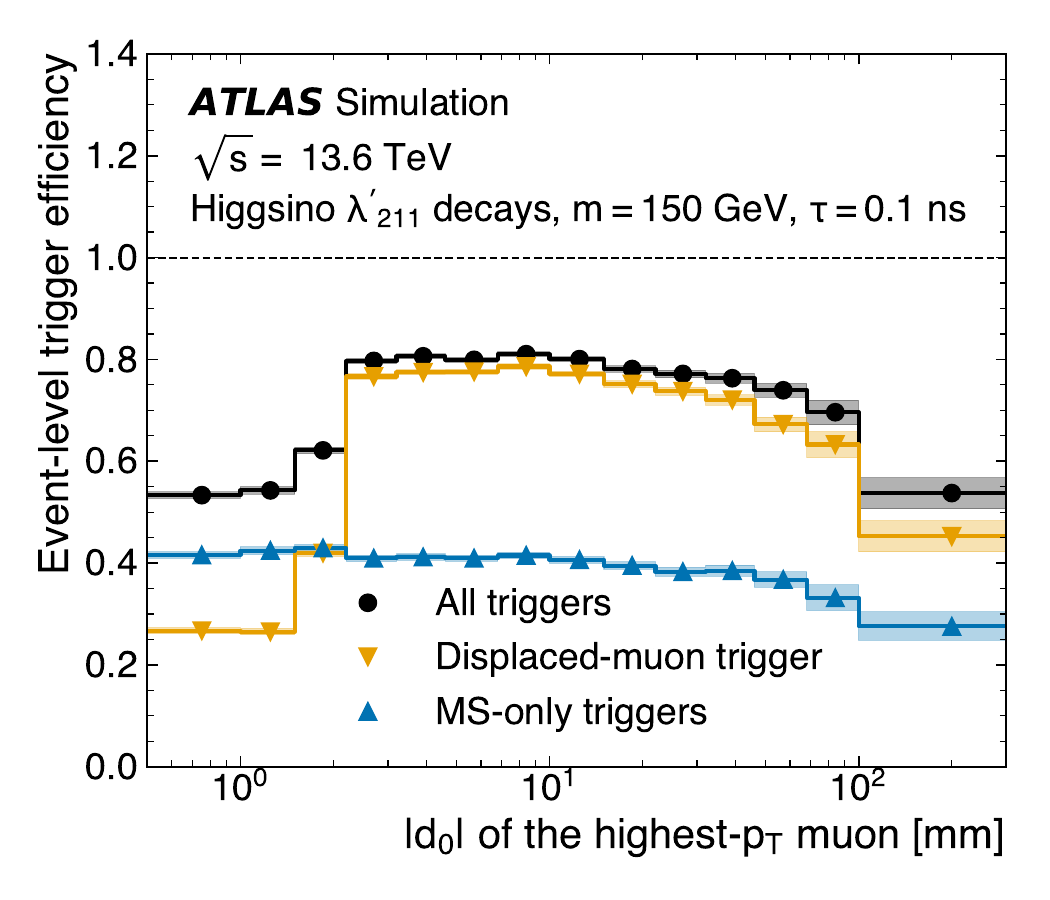}}
\end{minipage}
\hspace{0.1\linewidth}
\begin{minipage}{0.3\linewidth}
\centerline{\includegraphics[width=\linewidth]{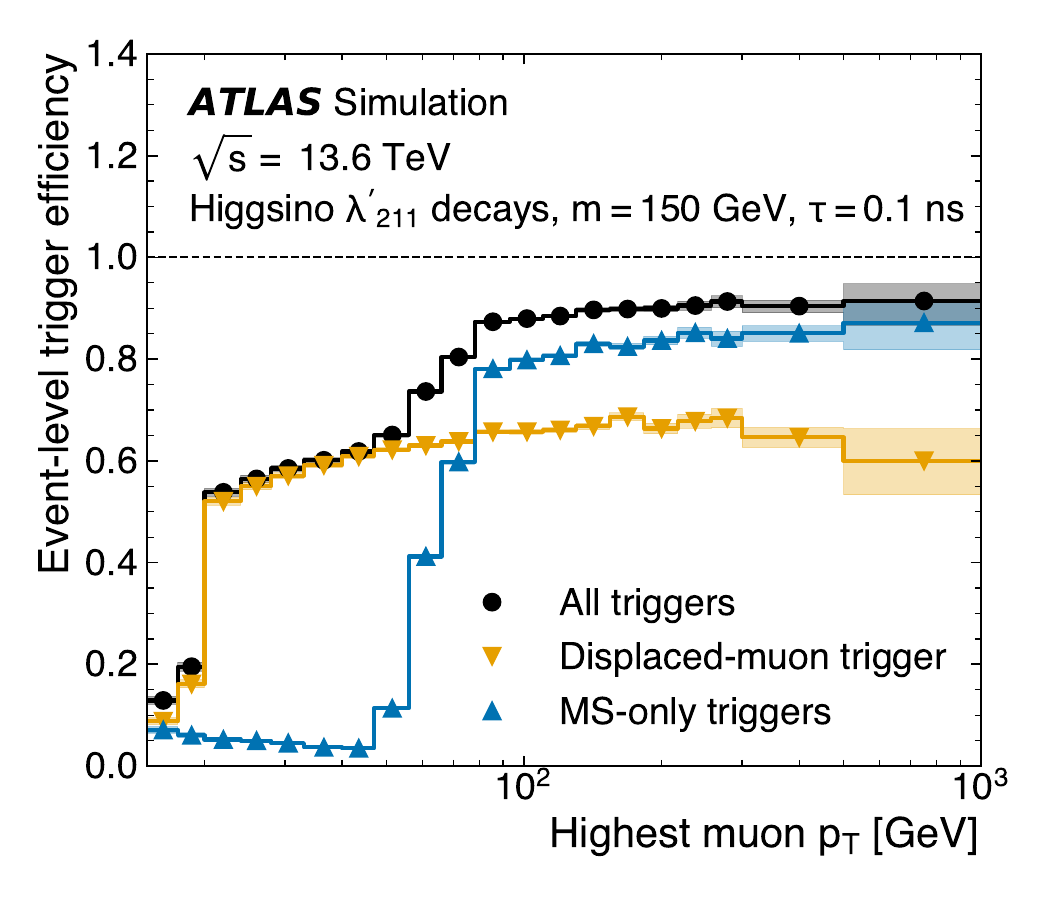}}
\end{minipage}
\caption[]{Trigger efficiency of the new displaced muon trigger compared to the regular muon triggers expressed in terms of muon displacement (left) and muon $p_{\textrm{T}}$ (right)\cite{DVMuon}.}
\label{fig:DVMuonTrig}
\end{figure}

This partial Run 3 (2022-2024, \SI{164}{fb^{-1}}) analysis\cite{DVMuon} is a follow-up of the full Run 2 (2016-2018, \SI{137}{fb^{-1}}) search\cite{DVMuonOLD}, and represents the first ATLAS LLP result using 2024 data. 
On top of the increase in data, the major improvement over the previous search is the introduction of a new displaced muon trigger based on an implementation of LRT at the trigger stage.
When used in addition to the normal muon spectrometer-only triggers, this allows for a noticeable increase in trigger efficiency for highly-displaced muons as seen in Figure~\ref{fig:DVMuonTrig} (left), and therefore better sensitivity to lower transverse momentum ($p_{\textrm{T}}$) muons as seen in Figure~\ref{fig:DVMuonTrig} (right).

\subsection{Selections}
The analysis is split into two SRs that focus on finding DVs that are very displaced ($\geq \SI{4}{mm}$, known as the "Far" SR), and those that are not as displaced ($\leq\SI{4}{mm}$ but $\geq\SI{1}{mm}$, known as the "Near" SR).
All DVs must have 4 or more tracks. 
The "Near" SR requires an DV invariant mass of $m_{\textrm{DV}}\geq \SI{40}{GeV}$, while the "Far" SR relaxes this requirement to $m_{\textrm{DV}}\geq \SI{20}{GeV}$.
All DVs must not be located inside detector material, and must lie within $|\textrm{DV}_{rxy}|\leq\SI{300}{mm}$ and $|\textrm{DV}_{z}|\leq\SI{300}{mm}$. 

As for the muons they must pass either one of two regular muon triggers or the displaced muon trigger.
The muons must have $p_{\textrm{T}}\geq\SI{25}{GeV}$, longitudinal and transverse impact parameters within $|z_0|<\SI{300}{mm}$ and $\SI{2}{mm}<|d_0|<\SI{300}{mm}$, and a transverse impact parameter significance of $|d_0|/\sigma_{d_0}>150$.
No requirements are imposed on whether the displaced muon originates from the DV.
Most importantly, muons must pass three vetoes specially designed to remove backgrounds from heavy-flavour, cosmic rays, and fakes.

\subsection{Background Estimate}
Processes that create background muons and background DVs are independent of each other. 
This means that the fraction of events that have background muons that pass the veto vs. get rejected by it is the same if one is looking at events with a "Near" DV, "Far" DV, or no DVs at all.
To simplify the analysis uses an ABCD method with these ratios to determine the number of events in the signal regions.
This is done separately for each muon background source.
Validation regions are created by inverting certain DV requirements. 

\subsection{Results}
The number of estimated background events in the different signal regions and the contributions from the different muon background sources, as well as the actual number of observed events is shown in Table~\ref{tab:DVMuonResult}. 
The dominant muon background is from heavy flavour processes. 
No significant excess was observed.

\begin{table}
\caption[]{Number of observed and expected events in the various signal regions, as well as the relative contributions of the different sources of uncertainties\cite{DVMuon}.}
\label{tab:DVMuonResult}
\vspace{0.4cm}
\begin{center}
\begin{tabular}{lcc}
\hline
Category        & "Near" DV SR   & "Far" DV SR   \\ \hline
Heavy Flavour   & $2.3\pm 0.8$    & $1.2\pm 1.0$   \\
Cosmic Rays     & $0.23\pm 0.13$  & $0.18\pm 0.10$ \\
Fakes in Barrel & $0.31\pm 0.15$  & $0.31\pm 0.15$ \\
Fakes in Endcap & $0.06 \pm 0.04$ & $0.11\pm 0.06$ \\
Total Estimated & $2.9\pm 0.8$    & $1.8\pm 1.1$   \\ \hline
Observed        & 1               & 3              \\ \hline
\end{tabular}%
\end{center}
\end{table}

\begin{figure}
\centering
\includegraphics[width=0.7\linewidth]{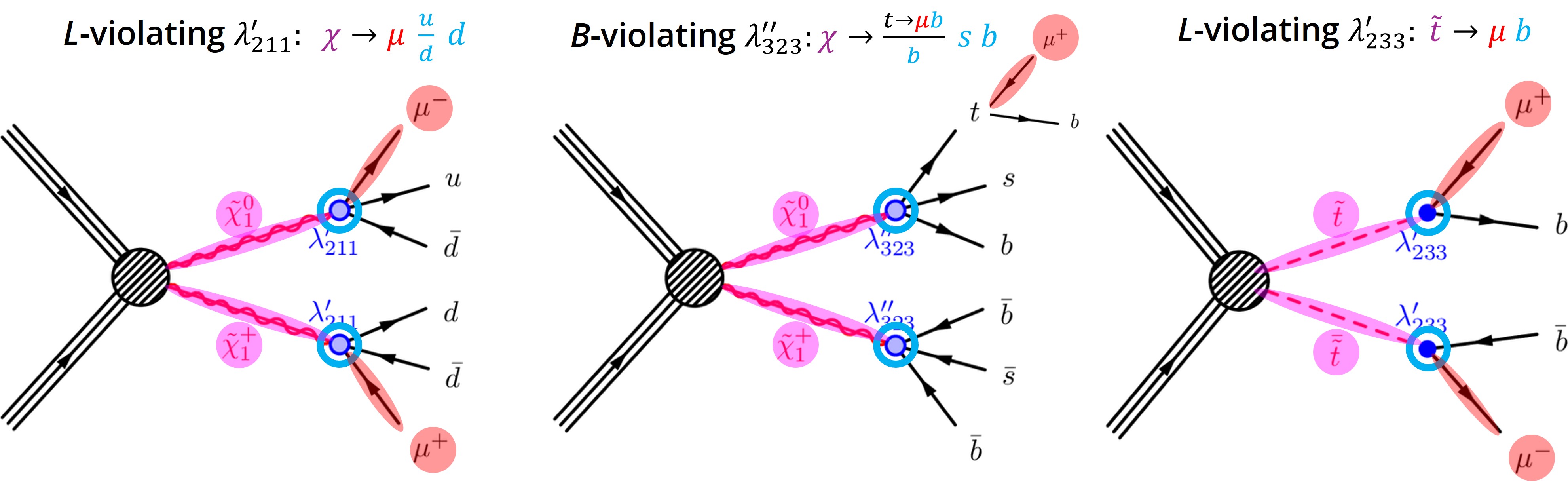}
\caption[]{Example Feynman diagrams of the RPV couplings used in the benchmark model\cite{DVMuon}. BSM particles are written in red. LLPs are highlighted in pink, the muon is highlighted in red, and the DV is circled in blue.}
\label{fig:DVMuonFeynman}
\end{figure}

$R$-parity violating (RPV) supersymmetry was used as the benchmark model, where the $\tilde{\chi}^0_1$, $\tilde{\chi}^\pm_1$, and $\tilde{\chi}^0_2$ are within \SI{1}{GeV} of each other and the other particles are decoupled.
Small RPV couplings make the $\tilde{\chi}$'s long-lived, and able to decay to fully Standard Model final states.
Three RPV couplings were considered as these result in final states with both a muon and quark content, as shown in Figure~\ref{fig:DVMuonFeynman}.
These are the $L$-violating $\lambda'_{211}$ (where $\tilde{\chi}\rightarrow \mu d (u/d)$), the $B$-violating $\lambda''_{323}$ (where $\tilde{\chi}\rightarrow sb(b/t)$ where $t\rightarrow \mu b$), and the $L$-violating $\lambda'_{233}$ (which allows $\tilde{t}\rightarrow \mu b$). 

\begin{figure}
\begin{minipage}{0.3\linewidth}
\centerline{\includegraphics[width=\linewidth]{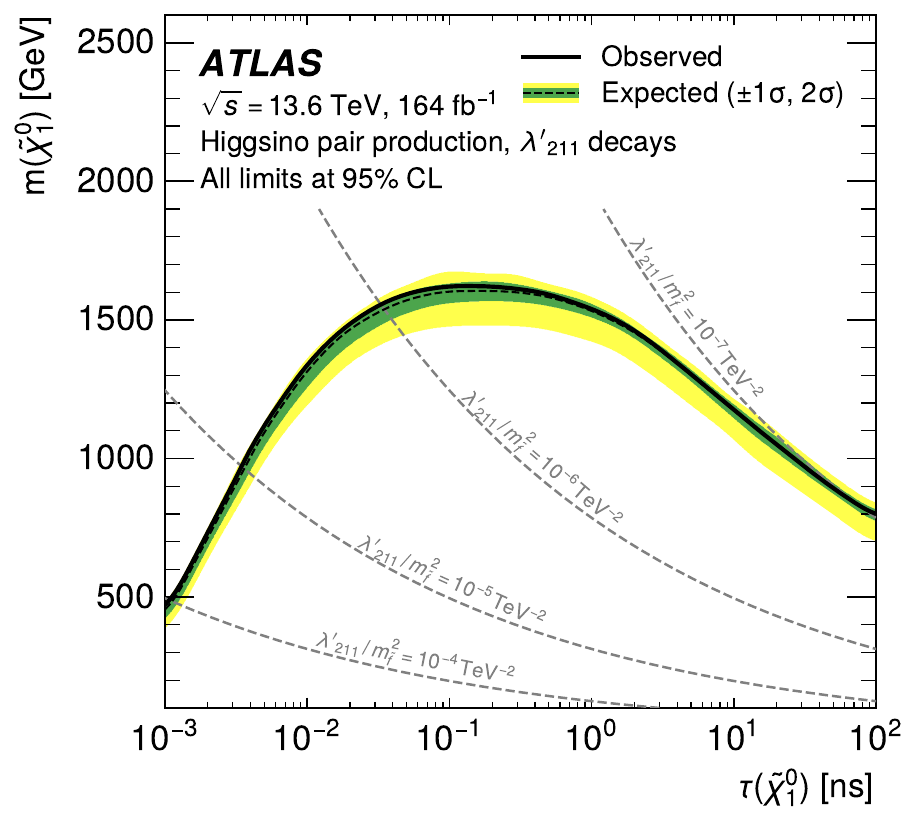}}
\end{minipage}
\hfill
\begin{minipage}{0.3\linewidth}
\centerline{\includegraphics[width=\linewidth]{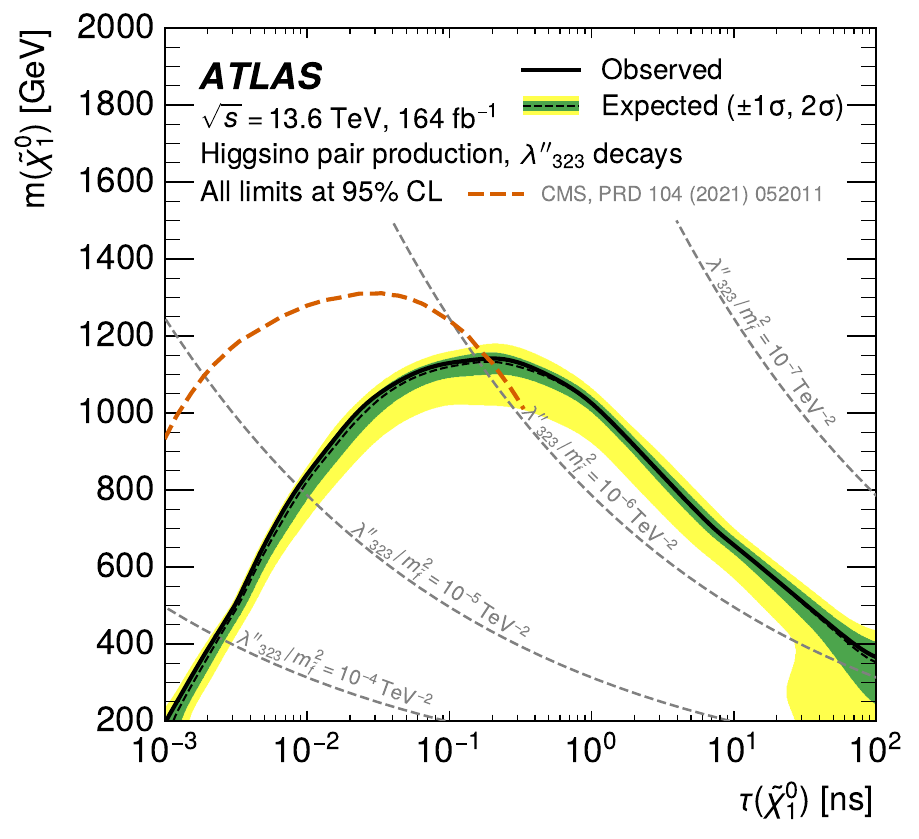}}
\end{minipage}
\hfill
\begin{minipage}{0.3\linewidth}
\centerline{\includegraphics[width=\linewidth]{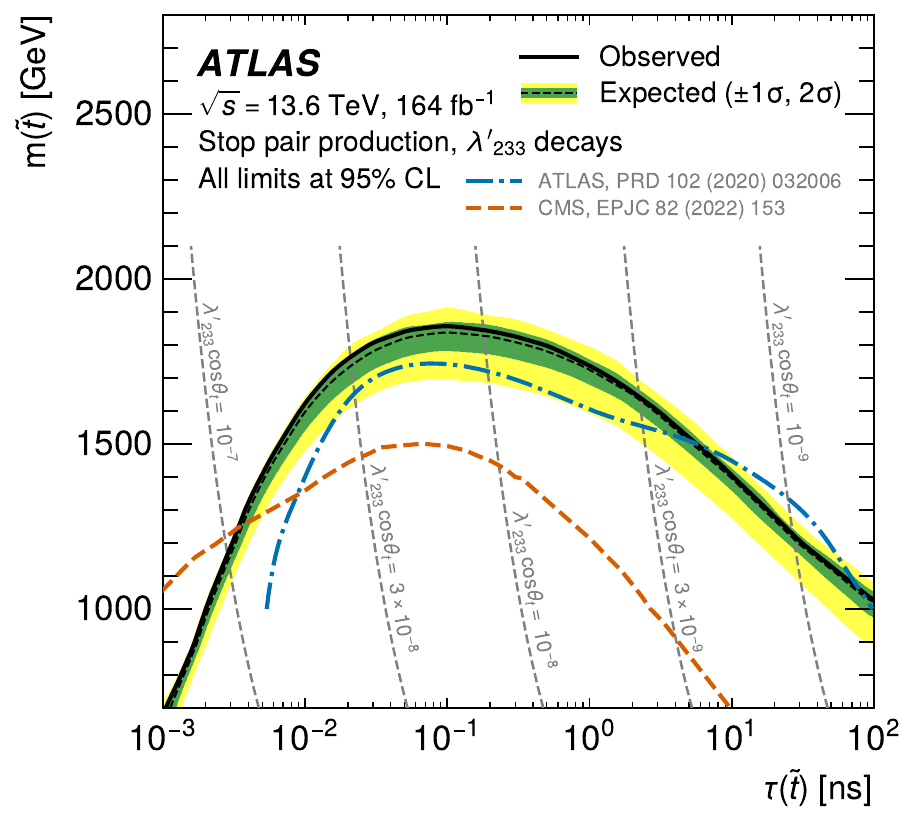}}
\end{minipage}
\caption[]{Exclusion limits for the $\lambda'_{211}$, $\lambda''_{323}$, and $\lambda'_{233}$ RPV couplings respectively\cite{DVMuon}.}
\label{fig:DVMuonLimits}
\end{figure}

The exclusion limits from each RPV coupling are shown in Figure~\ref{fig:DVMuonLimits}. 
The $\lambda'_{211}$ limits see an improvement over the ATLAS Run 1 higgsino-like neutralinos search\cite{ATLASRun1HiggsinoNeutralino}.
The $\lambda''_{323}$ limits are seen to be complementary to the CMS Run 2 hadronic DV pairs search, where they have optimized for pairs inside the beampipe and therefore better sensitivity at lower lifetimes\cite{CMSRun2HadronicDVPairs}.
Finally, the $\lambda_{233}$ limits see an improvement over the previous ATLAS Run 2 DV muon search\cite{DVMuonOLD}.

\section{Summary}
The ATLAS Collaboration continues to develop new and interesting strategies to increase sensitivity to long-lived particles.
From a new fuzzy vertexing algorithm to a new displaced muon trigger, the collaboration looks forward to the implementation of more long-lived particle searches in the future as the LHC soon enters its high luminosity phase.

\section*{References}
\bibliography{moriond}


\end{document}